\documentstyle[aps,prl,twocolumn,epsf]{revtex}
\begin{document}
%\flushbottom
\draft
\twocolumn[\hsize\textwidth\columnwidth\hsize\csname
@twocolumnfalse\endcsname
\title{Interband proximity effect and nodes of
superconducting gap in Sr$_2$RuO$_4$}
\author{M. E. Zhitomirsky$^{1,2}$ and T. M. Rice$^1$}
\address{
$^1$Institut f\"ur Theoretische Physik, ETH-H\"onggerberg, CH-8093 
Zurich, Switzerland \\
$^2$European Synchrotron Radiation Facility, B.P. 220, F-38043 
Grenoble, France}
\date{\today}
\maketitle
\begin{abstract}
\hspace*{2mm}
The power-law temperature dependences of the specific heat, the 
nuclear relaxation rate, and the thermal conductivity suggest 
the presence of line nodes in the superconducting gap of 
Sr$_2$RuO$_4$. These recent experimental observations contradict 
the scenario of a nodeless $(k_x+ik_y)$-type superconducting order 
parameter. We propose that interaction of superconducting order 
parameters on different sheets of the Fermi surface is a key to 
understanding the above discrepancy. A full gap exists in the 
active band, which drives the superconducting instability, while 
line nodes develop in passive bands by interband proximity effect.
\end{abstract}
\pacs{PACS numbers: 
             74.70.Pq,    %Ruthenates 
             74.20.Rp,    %Pairing symmetries (other than s-wave) 
             74.25.Bt     %Thermodynamic properties  
}
\vspace{5mm}
]
\narrowtext

The layered perovskite Sr$_2$RuO$_4$ with $T_c\simeq 1.5$~K
\cite{Maeno94} is an example of an unconventional superconductor with
non-$s$-wave Cooper pairing \cite{PT01}. The theoretical
proposal \cite{Rice95,Dan97,Sigrist99} of a spin-triplet $p$-wave
order parameter $\Delta_{\alpha\beta}({\bf k}) =
(i\sigma^y\sigma^z)_{\alpha\beta}\,d({\bf k})$, 
$d({\bf k})\propto(k_x+ik_y)$ is supported by experimental 
observations of a
temperature independent Knight shift for $H\perp c$ \cite{Ishida98}
and an increased muon spin-relaxation below $T_c$ \cite{Luke98}. Such
an axial gap function has a nonvanishing amplitude on the cylindrical
quasi-two-dimensional Fermi surface of Sr$_2$RuO$_4$
\cite{dHvA96,LDA,dHvA00}. This property favors the axial state as a
natural choice in a weak-coupling theory, which generally supports
nodeless solutions \cite{Rice95}. Recent experimental data collected
on high quality samples, however, seem to invalidate the above
conclusion. The power-law temperature dependences as $T\to 0$ found
for the specific heat, $C(T)\propto T^2$ \cite{NishiZaki99,NishiZaki00}, 
the NQR relaxation rate, $T_1^{-1}\propto T^3$ \cite{Ishida00},
the thermal conductivity, $\kappa(T)\propto T^2$ \cite{Tanatar00,Izawa}, 
the penetration depth \cite{Bonalde}, and the ultrasonic attenuation
\cite{Lupien} point to lines of zeros in the superconducting gap and, 
thus, question the consistency of the whole picture.

There have been several theoretical attempts to resolve this
controversy \cite{Miyake,Hasegawa,Graf,Maki}. Most suggest replacing
the axial $p$-wave order parameter $d_p({\bf k})\propto(k_x+ik_y)$ by
a suitable $f$-wave gap: $d_f({\bf k})\propto(k_x+ik_y) g({\bf k})$, 
where the even parity function $g({\bf k})$ is chosen to have
zeros, e.g.\ $k_xk_y$ or $(k_x^2-k_y^2)$ \cite{Hasegawa,Graf,Maki}. 
There is no clear microscopic mechanism for such an $f$-wave
instability. More importantly, nodes in an $f$-wave gap are only 
{\it marginally} stable, i.e. they disappear if all symmetry allowed
harmonics are included in the expansion of the gap function. For
example, for $g({\bf k})=k_xk_y$ one finds:
\begin{equation}
d ({\bf k}) = \eta_1 (k_x+ik_y)k_xk_y + i\eta_2 (k_x-ik_y) \ ,
\label{sym}
\end{equation}
where $\eta_1$ and $\eta_2$ are real. Both terms in Eq.~(\ref{sym})
transform in the same way under operations of the symmetry group of
the superconducting state. In particular, both harmonics are symmetric
with respect to a four-fold rotation $e^{i\pi/2}C_4$ and time-reversal
in combination with a reflection in the (100) plane,
$e^{i\pi}\hat{\cal T}\hat\sigma_x$. Therefore, the $f$- and the
$p$-wave harmonics are symmetry indistinguishable in tetragonal
crystals and mix with each other producing a finite gap $|\Delta|_{\rm
min}\sim\eta_2$.  Very recently, Izawa {\it et al\/}.\ have measured
the basal plane anisotropy of the thermal conductivity
$\kappa(\theta)$ in Sr$_2$RuO$_4$ at finite magnetic fields
\cite{Izawa}. Their results also discard an $f$-wave gap together with
a so-called `anisotropic' $p$-wave state \cite{Miyake} as possible
candidates to explain line of nodes in Sr$_2$RuO$_4$: all these
superconducting states have a substantial anisotropy of
$\kappa(\theta)$ in magnetic field determined by in-plane node
structure, whereas experimentally the basal plane anisotropy is much
smaller.  Izawa {\it et al\/}.\ \cite{Izawa} suggest instead horizontal lines of
nodes in the superconducting gap.

An opposite conclusion has been reached by Lupien {\it et al\/}.\ \cite{Lupien}
from the anisotropy of the ultrasonic absorption. However, the measured 
anisotropy appears mainly in the absolute magnitude of the attenuation
and not in the exponent of the temperature power
law. For this reason it is not clear that these results are in
conflict with horizontal lines of zeros and, as Lupien {\it et al\/}.\ stress,
detailed calculations based on the actual electronic structure are
necessary for a definitive interpretation of their results. 

Here, we propose a mechanism for the formation of horizontal line
nodes in the superconducting gap of Sr$_2$RuO$_4$. The Fermi energy
crosses three bands, determined by the $d_{xy}$ ($\gamma$-sheet of the
Fermi surface) and the hybridized $d_{xz}$ and $d_{yz}$ ($\alpha$- and
$\beta$-sheets) orbitals of Ru \cite{dHvA96,LDA,dHvA00}.  Magnetic
fluctuations, responsible for anisotropic Cooper pairing 
\cite{Rice95,Mazin}, have significant orbital dependence \cite{Imai}.  
Therefore, the intrinsic temperature of the 
superconducting instability should vary from band to band with one
sheet being the active source for superconducting instability and the
others being the passive sheets. In reality, interband scattering of
Cooper pairs, or proximity effect in the momentum space, will induce
the superconducting gap simultaneously on all parts of the Fermi
surface.  Such interband scattering is generally a strong effect,
which allows one to treat numerous multiband superconductors by an
effective single band Fermi surface.  Agterberg and co-workers
\cite{Dan97} have argued that Sr$_2$RuO$_4$ is different: a direct
in-plane scattering of the $p$-wave Cooper pairs between bands is
significantly suppressed by the orbital symmetry. Therefore they
conclude, one or two bands develop only tiny superconducting gaps,
which show up at intermediate temperatures as a residual density of
states with a subsequent crossover as $T\to 0$ to a full gap behavior.
In this Letter we study additional interlayer contributions to
interband scattering of Cooper pairs, which become important when
direct in-plane scattering is suppressed. We find that a nodeless
axial order parameter $d_p({\bf k})\propto(k_x+ik_y)$ in the active
band can induce superconducting gaps with zeros in the passive bands:
$d_p'({\bf k})\propto(k_x+ik_y)\cos(k_z/2)$. Thus, circular nodes of the
superconducting gap develop about the $c$ axis on one or two of the
three Fermi surface sheets. This model of weakly-coupled
superconducting order parameters in different bands fits well $C(T)$
in zero field \cite{NishiZaki99} and helps to explain the observed
field behavior of the specific heat \cite{NishiZaki00}.

We start with a general two-particle interaction:
\begin{equation}
\hat{V} = \int d{\bf r}d{\bf r}'U({\bf r},{\bf r}')
\psi^\dag({\bf r})\psi^\dag({\bf r}')\psi({\bf r}')\psi({\bf r}) \ ,
\end{equation}
where for simplicity we omit all spin indices assuming a fixed spin
structure of the triplet order parameter.  The band representation for
the effective interaction $\hat{V}$ is obtained by (i) expanding the
field operators $\psi({\bf r})$ in terms of the band operators:
$\psi({\bf r})=\sum_{l,{\bf k}}\varphi_{l{\bf k}}({\bf r}) c_{l{\bf k}}$ 
($l$ is a band index) and (ii) representing the Bloch function
$\varphi_{l{\bf k}}({\bf r})$ in a given band as a lattice sum over the
Wannier function of the Ru orbitals: $\varphi_{l{\bf k}}({\bf r})=
\sum_n e^{i{\bf k}{\bf R_n}} \phi_l({\bf r}-{\bf R}_n)$. The interaction in
the Cooper channel is
\begin{equation}
\hat{V} = \frac{1}{2} \sum_{ll',{\bf k}{\bf k}'}
V_{ll'}({\bf k},{\bf k}') c^\dag_{l{\bf k}}
c^\dag_{l-{\bf k}} c_{l'-{\bf k}'}c_{l'{\bf k}'}  \ ,
\end{equation}
where the scattering vertex is given by
\begin{eqnarray}
&&V_{ll'}({\bf k},{\bf k}')\!=\!\!\int\!\! 
d{\bf r}d{\bf r}' U({\bf r},{\bf r}')\!\!\!\sum_{nn'mm'} \!\!\!
e^{-i{\bf k}({\bf R}_n\!-\!{\bf R}_{n'})}
e^{i{\bf k}'({\bf R}_m\!-\!{\bf R}_{m'})} \nonumber \\
&&\times\phi^*_l({\bf r}-{\bf R}_n)\phi^*_l({\bf r}'-{\bf R}_{n'})
\phi_{l'}({\bf r}'-{\bf R}_{m'})\phi_{l'}({\bf r}-{\bf R}_m) .
\label{sum}
\end{eqnarray}
Following Ref.~\cite{Dan97} we now consider interband scattering
processes ($l\neq l'$) in the tight-binding approximation, i.e.\ 
assume that the Wannier functions are well localized and, therefore,
the main contribution to $V_{ll'}({\bf k},{\bf k}')$ comes from a few
neighboring sites. The largest on-site contribution (${\bf R}_n\!=\!
{\bf R}_{n'}\!=\!{\bf R}_m\!=\!{\bf R}_{m'}$) is
independent of $\bf k$ and ${\bf k}'$.  It causes coupling only
between conventional $s$-wave order parameters in two bands.  The
coupling of the $p$-wave order parameters appears first in the sum
(\ref{sum}) for ${\bf R}_m\!=\!{\bf R}_n$, ${\bf R}_{m'}\!=\!{\bf R}_{n'}$, 
and $({\bf R}_{n'}-{\bf R}_n)=\mbox{\boldmath$\delta$}_i=
\pm a\hat{\bf x}(\hat{\bf y})$ ($a$ is lattice constant):
\begin{eqnarray}
&&V^{pp}_{ll'}({\bf k},{\bf k}')=\sum_{i=x,y}\sin k_ia\sin
k'_ia\nonumber\\
&&\mbox{}\times \int d{\bf r}d{\bf r}' U({\bf r},{\bf r}')
\phi^*_l({\bf r})\phi_{l'}({\bf r})
\phi^*_l({\bf r}' -\mbox{\boldmath$\delta$}_i)
\phi_{l'}({\bf r}'-\mbox{\boldmath$\delta$}_i) .
\label{direct}
\end{eqnarray}
This direct in-plane scattering of the Cooper pairs induces
the same nodeless superconducting gap
\begin{equation}
d_1({\bf k})\propto (\sin k_xa + i\sin k_ya)
\label{typeI}
\end{equation}
on all sheets of the Fermi surface. If we now approximate in a
tight-binding spirit $U({\bf r},{\bf r}')\approx
U(\mbox{\boldmath$\delta$}_i)$, then the double integral in
Eq.~(\ref{direct}) factorizes in a product of two spatial integrals.
Each of these integrals vanishes separately for $l=\gamma$ and
$l'=\alpha,\beta$ because the orbitals from these bands have different
parity with respect to $\hat\sigma_z$. Thus, it is essential to keep
spatial dependence of $U({\bf r},{\bf r}')$. For a Coulomb-type
interaction the off-diagonal matrix element ($l\neq l'$) in
Eq.~(\ref{direct}) has a dipolar reduction $(b/a)^2\simeq 0.02$
\cite{Dan97} compared to the diagonal matrix elements ($l=l'$), where
$b$ is a characteristic spatial extension of the Wannier functions.
In reality, the matrix element will be reduced even further because
$d_{xz}$ and $d_{yz}$ orbitals mix in $\alpha$- and $\beta$-bands only
in the close vicinity of the Brillouin zone diagonals. Away
from these directions there is an extra approximate symmetry
$\hat{\sigma}_{x(y)}$, which introduces an effective quadrupolar
reduction of the matrix element in Eq.~(\ref{direct}).
Thus, the direct interaction of the $p$-wave order parameters between
$\gamma$- and $\alpha$- or $\beta$-bands is significantly reduced and
the amplitude of the type-I gap (\ref{typeI}) in passive bands is much
smaller than in the active band \cite{MS}. 

The next contribution to the interband scattering of the $p$-wave
pairs in Eq.~(\ref{sum}) comes from interlayer terms with ${\bf R}_m\!=\! 
{\bf R}_n$ and ${\bf R}_{n'}\!=\!{\bf R}_n\pm a \hat{\bf x} 
(\hat{\bf y})$, ${\bf R}_{m'}\!=\!{\bf R}_n\pm \frac{a}{2}\hat{\bf x}\pm
\frac{a}{2} \hat{\bf y}\pm \frac{c}{2}\hat{\bf z}$ on
a bct lattice of Ru-atoms.  Summing over all contributing sites the
$p$-wave gap of the type (\ref{typeI}) in the $\gamma$-band induces
\begin{equation}
d_2({\bf k})\propto\left(\sin\frac{k_xa}{2}\cos\frac{k_ya}{2}
+ i \sin\frac{k_ya}{2}\cos\frac{k_xa}{2}\right)
\cos\frac{k_zc}{2}
\label{typeII}
\end{equation}
in $\alpha$- and $\beta$-bands and vice versa. Existence of the
type-II $p$-wave gap, but in all bands simultaneously, has been
conjectured by Hasegawa {\it et al\/}.\ \cite{Hasegawa}. They, however,
based their suggestion on an (unjustified) assumption of a repulsive
interaction between electrons in a single Ru-O plane and an attraction
only for electrons in adjacent layers.  The type-II superconducting
gap $d_2({\bf k})$ has circular line nodes at $k_z=\pm\pi/c$. Importantly,
they are stable with respect to an admixture of a small amount of the
type-I gap, which only shifts the position of zeros along $k_z$. The
two gaps are mixed with a real phase, as required by the time-reversal
symmetry ($\hat{\cal T}\hat\sigma_x$), and nodes of
$d_1+d_2$ disappear only for $|d_1|_{\rm
  max}>|d_2|_{\rm max}$.

A reliable estimate for the strength of the scattering vertices
corresponding to the two types of induced gaps (\ref{typeI}) and
(\ref{typeII}) must be done in a truly microscopic treatment of the
Fermi liquid state in Sr$_2$RuO$_4$, which is beyond the scope of our
work. Note that scattering processes contributing to
Eq.~(\ref{typeII}) have one symmetry cancellation factor less than
Eq.~(\ref{typeI}), but instead they are reduced by the small overlap
of the orbitals in adjacent layers. Information on interlayer overlap
can be obtained by analyzing the results of the high-precision de
Haas-van Alphen measurements \cite{dHvA00}, which determined
corrugation of the Fermi surface cylinders along $k_z$.  Bergemann
{\it et al\/}.\ \cite{dHvA00} found the strongest corrugation in the
$\beta$-sheet of the Fermi surface: $\Delta
k_{F,\beta}\sim\cos(k_zc/2)$ and a much weaker corrugation of the
$\gamma$-sheet: $\Delta k_{F,\gamma}\sim\cos(k_zc)$. This different
periodicity naturally appears in a tight-binding model, since $d_{xz}$
($d_{yz}$) orbitals have a direct interlayer overlap $t'_\perp$,
leading to a diagonal contribution to the band energy:
$8t'_\perp\cos(k_xa/2)\cos(k_ya/2)\cos(k_zc/2)$. On the other hand,
planar $d_{xy}$ orbitals do not hybridize across the layers.  However,
they can hybridize with $d_{xz}$ ($d_{yz}$) orbitals in adjacent
planes with the matrix element $t''_\perp$, which contributes
$8t''_\perp\cos(k_{x(y)}a/2)\sin(k_{y(x)}a/2)\sin(k_zc/2)$ to the
off-diagonal kinetic energy. As a result, the corrugation of the
$\beta$-cylinder is a first-order effect in $t'_\perp$, whereas the
corrugation of the $\gamma$-sheet has a much weaker second-order
contribution $\sim[t''_\perp\cos(k_zc/2)]^2/t$.  Comparing the
experimental corrugations we find: $t'_\perp\simeq -1$~meV and
somewhat larger magnitude for $t''_\perp\simeq 3$~meV.  Since the
interlayer hopping amplitude is proportional to the orbital overlap
in Eq.~(\ref{sum}), we argue that it is quite possible to have a
comparable amplitude for the two types of the $p$-wave gaps induced by
interband proximity effect in passive bands.

We consider now the effect of line nodes in passive bands on
thermodynamic properties of the superconducting state. In particular,
we calculate the specific heat $C(T)$. Since the Cooper pair
scattering between the $\alpha$- and the $\beta$-sheets is not small,
we adopt an effective two band model for Sr$_2$RuO$_4$ 
and split the total density of states at the Fermi level according to 
$N_{01}:N_{02}=N_{0\gamma}:(N_{0\alpha}+N_{0\beta}) = 0.57:0.43$ 
based on the de Haas-van Alphen measurements \cite{dHvA96}. We
also assume that the active sheet for the superconducting instability
is the $\gamma$-sheet.  Our main motivation for this assumption comes
from comparison with the experimental data below. We adopt a
weak-coupling approach and parameterize the pairing potential in the
two bands by three parameters: $V_{11}({\bf k},{\bf k}')=-g_1f({\bf k}) 
f({\bf k}')$, $V_{22}({\bf k},{\bf k}')=-g_2\tilde{f}({\bf k})
\tilde{f}({\bf k}')$, and $V_{12}({\bf k},{\bf k}')=-g_3 f({\bf k}) 
\tilde{f}({\bf k}')$, where we choose for simplicity $f({\bf k})= 
{\bf k}/k_F$ and $\tilde{f}({\bf k})=\sqrt{2}({\bf k}/k_F) \cos(k_z/2)$, 
i.e. we presume that only interlayer processes
contribute to the interband scattering of the Cooper pairs.  The
pairing interaction in the active band is 
attractive ($g_1>0$), while interaction constants in the passive band
($g_2$) and between the bands ($g_3$) can have arbitrary sign. Solving
the system of the two gap equations numerically we determine the
specific heat from
\begin{equation}
C(T) = 2 \sum_{l,\bf k} E_{l\bf k} \frac{d f(E_{l\bf k})}{dT} \ ,
\end{equation}
where $E_{l\bf k}$ is the quasiparticle excitation energy
$\sqrt{\epsilon_{l\bf k}^2 + \Delta^2_l({\bf k})}$ (we consider only
unitary triplet states) and $f(E_{l\bf k})$ is the corresponding Fermi
distribution.
\begin{figure}
\unitlength1cm
\begin{picture}(2,7.5)
\epsfxsize=7.7cm
\put(-0.3,-2.3){\epsffile{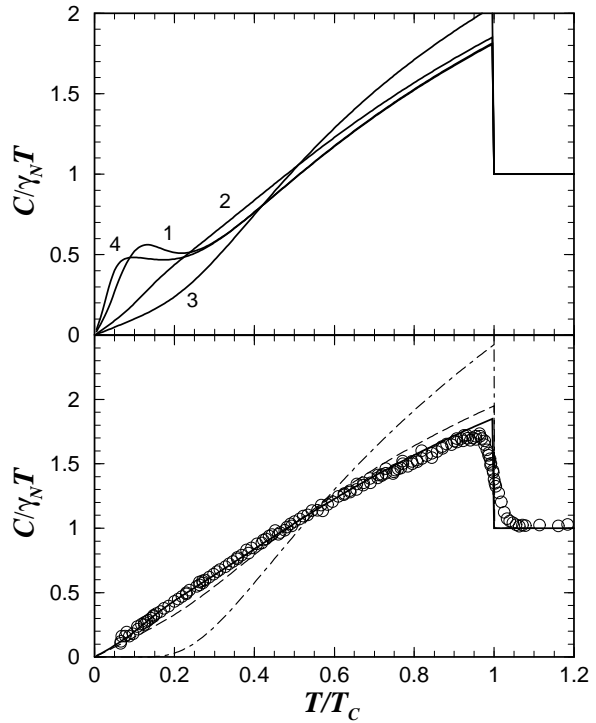}}
\end{picture}
\vspace{2.5cm}
\caption{Temperature dependence of the normalized specific heat.
The upper panel: the two-band model results
for various choices of the interaction parameters.
Curves \#1--3 correspond to $g_2/g_1=0.85$ and
$g_3/g_1=0.01,0.07,0.2$, respectively.
The curve \#4 is for $g_2/g_1=0.1$ and $g_3/g_1=0.07$.
The lower panel:
circles are experimental data for Sr$_2$RuO$_4$ [11].
One-band results are shown for anisotropic gap 
with line nodes (dashed line) and for isotropic gap 
(dot-dashed line). Solid line is the two-band model fit 
with $g_2=0.85g_1$ and $g_3=0.07g_1$.}
\end{figure}

We present in the upper panel of Fig.~1 the specific
heat for different choices of the coupling constants
in the two-band model. The calculations 
have been done for a `typical' weak-coupling magnitude 
of $g_1=0.4$ (in units of the inverse total
density of states) varying the two other parameters. 
The first three curves correspond to weak $g_3=0.01g_1$ (\#1),
intermediate $g_3=0.07g_1$ (\#2), and strong 
$g_3=0.2g_1$ (\#3) interband scattering of the Cooper pairs,
keeping in all cases $g_2=0.85g_1$. 
Such a moderate change in the coupling constants between the two bands
$\lambda_2/\lambda_1=g_2N_{02}/g_1N_{01}= 0.64$ results in an order 
of magnitude difference in their bare transition temperatures: 
$T_{c2}/T_{c1}=0.086$ in the weak-coupling theory. 
For the weaker interband coupling the heat capacity develops a second
peak, which reflects a nonzero bare transition temperature
in the passive band. 
For the stronger interband coupling the two gaps are tightly 
bound to each other and we return to an effective one-band behavior.
The curve \#4 is an example of a shoulder in 
the temperature dependence of $C/T$, which arises for a reduced pairing 
interaction in the passive band $g_2=0.1g_1$ when we keep
the same scattering vertex $g_3=0.07g_1$ as for the curve \#2.

In the lower panel of Fig.~1 we present our fit to
the experimental data of NishiZaki {\it et al\/}.~\cite{NishiZaki00},
which coincides with the curve \#2 above, and also show the 
results of one-band models with constant and anisotropic gaps. Under 
condition of a reduced interband coupling the heat capacity
jump at $T_c$ is entirely determined by the active band.
Remarkable agreement between the experimental magnitude of the 
jump and $\Delta C/C_n\simeq 1.43 N_\gamma/N_0=0.82$, which is
obtained from the standard BCS result by its renormalization on 
the partial density of states of the $\gamma$-sheet, is a direct 
confirmation of our choice of the active band.
Though, it is impossible to fix all three 
parameters of the two-band model uniquely,
this model can naturally explain a clear convex shape 
of the experimental data for $C/T$ at low temperatures by 
choosing an intermediate
strength of the interband scattering matrix element. 
In contrast, a one-band model with anisotropic gap
or the two-band model with a strong
interband scattering
predict a concave shape for $C/T$.
From our fit we cannot also exclude a possibility
of a small but finite $|\Delta_{\rm  min}|$ in
the passive bands, which appears if in-plane scattering
amplitude slightly exceeds the interplane contribution
(\ref{typeII}).

The field dependence of the residual density of states at
low temperatures suggests
another argument in favor of the multiband scenario.  
Small magnetic fields ($\ll H_{c2}$)
quickly restore about 40\% of the total density of states
\cite{NishiZaki00}. In our model this corresponds to the behavior
exhibited by the curve \#4, the upper panel in Fig.~1, 
in the case of temperature effects.
Such a new feature in external field arises because of an additional
suppression of superconductivity in the passive $\alpha$- and
$\beta$-bands for $H\perp c$. Stronger $c$-axis dispersion in these bands 
leads to a larger coherence length $\xi_c$ and an extra reduction
of the bare $H_{c2}^{ab}$.
This effect should disappear for $H\parallel c$ because of 
similar values of the in-plane Fermi velocities, 
which also agrees with the experiment \cite{NishiZaki00}.
It would be also interesting to reinvestigate
the impurity effect on the residual density of states.
Such an analysis has been done previously for the two-band
model with a constant gap amplitude in the passive band \cite{Dan99}. 
Line nodes can modify the expected behavior and produce a gapless 
superconducting state in the passive bands.

In summary, we have shown that circular horizontal
line nodes in the superconducting
gap of Sr$_2$RuO$_4$ appear due to weak and anisotropic 
interband proximity effect. This effect is a consequence of 
(i) non $s$-wave symmetry of the Cooper pairs [conventional
superconductors have generally a strong isotropic interband coupling
dominated by on-site term in Eq.~(\ref{sum})] and (ii) specific
symmetry of the Ru orbitals, which give extra suppression of the matrix 
element in Eq.~(\ref{direct}). Further experimental tests of our
scenario should include studying effects of pressure, which can modify
the strength of interlayer scattering amplitude for the Cooper pairs. 

We thank E. M. Forgan, I. I. Mazin, M. Sigrist and L. Taillefer 
for stimulating discussions. This work has been supported by
Swiss National Fund.


\begin{thebibliography}{10}
\bibitem{MS}
Mazin and Singh claimed on the basis of their LDA-type
calculations that there is no such a reduction of interband vertices
\cite{Mazin}. However, they implicitly assumed in their calculations
that the matrix element in Eq.~(\ref{direct}) is the same
for $l=l'$ and $l\neq l'$, which is incorrect.
\end{thebibliography}
\end{document}